\documentclass[12pt]{spieman}  
\usepackage{amsmath,amsfonts,amssymb}
\usepackage{graphicx}
\usepackage{setspace}
\usepackage{tocloft}
\usepackage{lineno}
\title{Effect of total dose proton irradiation on the performance of Kinetic Inductance Detectors for operation in a space based far-Infrared observatory}

\author[a]{Anaïs Besnard}
\author[b]{Giulia Conenna}
\author[a]{Valentin Sauvage}
\author[b]{Kenichi Karatsu}
\author[b]{Stephen J.C. Yates}
\author[b]{Jochem J.A. Baselmans}
\author[a]{Bruno Maffei}
\author[b]{Lorenza Ferrari}
\affil[a]{IAS, Institut d’Astrophysique Spatiale, Université Paris-Saclay, CNRS, Orsay, France}
\affil[b]{SRON, Space Research Organisation Netherlands, Niels Bohrweg 4, 2333 CA Leiden, and Landleven 12, 9747 AD Groningen, The Netherlands}

\cftpagenumbersoff{figure}
\cftpagenumbersoff{table} 
\begin{document} 
\maketitle

\begin{abstract}
Kinetic Inductance Detectors (KIDs) are a promising technology for future space missions, where exposure to high-energy particles may damage the detectors and degrade their performance. In this work, we discuss the effect of high energy proton beam irradiation on two different aluminum KID arrays, one based upon absorbers on 350~$\mu$m thick solid silicon (Si) and one based upon antennas on a 100~nm thick SiN membrane.
We used a total dose of 5.7~krad, equivalent to approximately 10 years of operation at the L2 Lagrange point. 
A comparison between pre-radiation and post-radiation measurements was done while keeping the detectors at 120 mK to prevent thermal annealing of possible defects created during proton beam irradiation. We find that there is no significant change in the quasi-particle lifetime $\tau_{qp}^*$ and the dark responsivity $d\theta/dP_{dark}$, but we do observe an increase in the noise and NEP that is attributed to limitations in the post radiation data analysis. 
\end{abstract}


{\noindent \footnotesize\textbf{*}Anaïs Besnard, \linkable{anais.besnard@universite-paris-saclay.fr} }

\begin{spacing}{2}   

\section{Introduction}
\label{sect:intro}  
A cold telescope in space, with its optics cooled to temperatures below 4.5 K, could provide orders-of-magnitude improvements in far-infrared observing speed compared with previous missions, such as Herschel/PACS~\cite{} and SPIRE~\cite{}. 
PRIMA (PRobe far-Infrared Mission for Astrophysics) has been proposed to exploit this capability. 
The mission concept includes two instruments, PRIMAger and FIRESS~\cite{PRIMA}, and aims to address three primary science goals: tracing the astrochemical signatures of planet formation, measuring the formation and buildup of galaxies, heavy elements, and dust, and probing the coevolution of galaxies and their supermassive black holes across cosmic time.

The detectors for PRIMA are Kinetic Inductance Detectors (KIDs), which are superconducting resonators whose resonant frequencies depend on the density of Cooper pairs and quasiparticles~\cite{PDay_2003}. When incoming radiation with a frequency exceeding the gap frequency (90~GHz for aluminum (Al) KIDs) is absorbed in the sensitive part of the detector, it breaks Cooper pairs into quasiparticle excitations, causing a measurable shift in the resonant frequency ($f_r$) and its Quality factor ($Q$). This response is then read out using homodyne detection. The sensitivity of state-of-the-art KID reaches a Noise Equivalent Power (NEP) of $\sim 10^{-20}$~W/$\sqrt{\rm Hz}~$\cite{Baselmans_2022, PDay2024}.
With its frequency-domain multiplexing (FDM) capability, allowing scaling to large array sizes, KIDs are one of the most attractive detector technologies for space missions in the wavelength from millimeter-waves to far-infrared.

The PRIMA satellite is supposed to be operated at the Lagrange point L2.
In space, hits from cosmic ray particles~\cite{Planck_2011} could damage detectors by ionization or knocked-on of atoms in the detector materials through elastic/inelastic scattering (e.g. Coulomb elastic scattering and nuclear elastic/inelastic scattering).
These damages could lead to deterioration of detector sensitivity during operation.
For example, cosmic ray interactions with the aluminum could reduce the quasi-particle lifetime ($\tau_{qp}^*$) due to increased disorder and trapping effects~\cite{Barends2009a, deRooij2025b} resulting in a reduction of responsivity and increase in NEP. Furthermore, the $Q$ factor could be degraded due to additional losses hence also leading to reduction in responsivity and increase in NEP (see equations in Sec.~\ref{KIDNEP}).

In a previous irradiation experiment described in Ref.~\citeonline{Karatsu_2015}, KIDs were placed in front of a proton beam (160~MeV) at room temperature, and no significant degradation in NEP was observed after about 10~krad dose.
However, since defects created in aluminum by particle hits are most likely mobile at room temperature, there is a concern that this experiment does not correctly represent the situation during flight~\cite{watkins_2000}.
In this paper, we present results of a cryogenic proton beam exposure, where the pre- and post exposure performance evaluation was done while keeping the detectors at 120 mK, the operational temperature of the instrument.

\section{Methods}

\subsection{Particle accelerator}
We have illuminated two KID arrays in front of the proton beam line of the Particle Therapy Research Center (PARTREC), located in Groningen University, the Netherlands~\cite{gerbershagen:hiat2022-mo3c2}. 
The cyclotron in PARTREC can deliver beams of various ions (protons to oxygen) with energies up to 190~MeV. For this experiment, we used a proton beam with an energy of 184~MeV.

\subsection{DRACuLA}
During the irradiation experiment, the KID arrays were cooled down to $120$~mK by the DRACuLA facility (Detector irRAdiation Cryogenic faciLity for Astrophysics). DRACuLA is a cryogenic setup developed to study particle impacts on new detector technologies at their operating temperature (50~mK to 800~mK)~\cite{dracula}. This dilution refrigerator (model LD400 from Bluefors) was designed with special features such as KF-50 ports and windows on each thermal shield to connect to a beam line. DRACuLA can be adaptable to any kind of particle accelerator and allows the characterization of detectors/focal planes' particle hits susceptibility for future space/balloon-borne missions. It can provide 500~$\mu$W of cooling power at 100~mK~\cite{besnard_cryogenic_2024}. 

The temperature stability of the coldest stage was monitored by a RuOx thermometer. A 120~$\Omega$ heater was also installed to maintain the KID arrays at their operating temperature (120~mK) using a PID. Both of them were handled by a Lakeshore~372 AC resistance bridge. To ensure the 120~mK during the beam irradiation, stability tests were performed on the cryostat before irradiation to verify that the cooling power was enough to handle the power of $\approx 300~\mu \text{W}$, which was the expected heat produced by the proton beam irradiation. Moreover, the PID parameters were tuned to ensure thermal stability at the detector level, allowing a $120~\text{mK} \pm 695\text{ $\mu$K}$ stability when the beam was on and $120~\text{mK} \pm 33\text{ $\mu$K}$ while the beam was off. The thermal performances during the irradiation campaign, together with the installation of DRACuLA are detailed in Ref.~\citeonline{Sauvage_prima}.

\subsection{KID arrays}
The PRIMAger instrument is composed of two focal planes: a hyperspectral focal plane consisting of absorber-coupled KIDs fabricated on a solid Si wafer~\cite{dabironezare2025},  and a polarimetric focal plane composed of antenna-coupled KIDs~\cite{Baselmans_2022}, with the aluminum section fabricated on a SiN membrane. Small representative prototypes of both type of KIDs were designed, fabricated, and characterized in a laboratory with adiabatic demagnetization refrigerator (ADR) before the irradiation campaign. The antenna-coupled KID array has 27 detectors and is the same as the one described in Ref.~\citeonline{Baselmans_2022}, while the absorber array has 30 KIDs with a design similar to the one described in Ref.~\citeonline{dabironezare2025}.

The antenna-coupled KIDs consist of a wide coplanar waveguide (CPW) section made of niobium-titanium-nitride (NbTiN) loaded with an interdigitated capacitor (IDC), fabricated on a silicon substrate (350~$\mu$m thick FZ Si, $\rho = 10$~k$\Omega\cdot\text{cm}$). At its shorted end the CPW is made of 16~nm aluminum with a 400~nm wide central line deposited on a 100~nm thick silicon-nitride (SiN) membrane. A combination of a spacer wafer, mounted using Perminex, and a lens array to provide efficient radiation coupling was used. The lens array was mounted on the spacer wafer with Epotek 2216. The prototype is presented in Figure~\ref{fig:setup} (c), with further details given in Ref.~\citeonline{Baselmans_2022}.

The absorber-coupled KIDs are fully fabricated on a silicon substrate (350~$\mu$m thick FZ Si, $\rho = 10$~k$\Omega\cdot\text{cm}$), and consist of a wide NbTiN CPW coupled at its shorted end via the absorber. The absorber is fabricated from aluminum line that is 300~nm wide and 21~nm thick. A lens array was mounted on the backside of the chip using Epotek 2216 epoxy glue. The prototype is presented in Figure~\ref{fig:setup} (b), with further details given in Ref.~\citeonline{dabironezare2025}.


\subsection{KID sensitivity}
\label{KIDNEP}

The KID dark (electrical) NEP is an estimate of the limiting performance of the detector at $T~<~T_c/8$ and negligible absorbed power. It can be obtained using \cite{janssen_equivalence_2014} :
\begin{equation}
    NEP_{\textrm{dark}}=\frac{\sqrt{S_{\theta}}}{d\theta /dP_{dark}},
    \label{Eqn:NEP}
\end{equation}
with $S_{\theta}$ representing the phase noise power spectral density (PSD), and $d\theta/dP_{dark}$ being the dark phase responsivity, which can be written as:
\begin{equation}
    \left(\frac{d\theta}{dP_{dark}}\right)=\frac{\eta_{pb}\tau_{qp}^*}{\Delta}\frac{d\theta}{dN_{qp}},
    \label{Eqn:response}
\end{equation} 
where the last term, $d\theta/dN_{qp}$, is the phase response per quasiparticle which depends on the aluminum material, and is also proportional to the resonator $Q$ factor. $\Delta$ is the energy gap of aluminum, $\eta_{pb}$ is the pair breaking efficiency~\cite{Guruswamy_2014}, and $\tau^*_{qp}$ is the experimental quasiparticle recombination time at $T < T_c/8$ where thermal effects are negligible and $\tau_{qp}^*$ represents the limiting (maximum) lifetime.
The quasiparticle number, $N_{qp}$, is given by $N_{qp}=n_{qp}V$ where $n_{qp}$ and $V$ represent the quasiparticle number density and the film volume, respectively.
$n_{qp}$ can be calculated from a bath temperature of KIDs, as described in Eq.~2.4 of Ref.~\citeonline{deVisserThesis2014}, with the $\Delta$ value determined from the measured critical temperature ($T_c$) of the film. $V$ is determined by the line width measured using a scanning electron microscope (SEM) and the line thickness measured using a profilometer.

Cosmic ray interactions with KIDs could degrade $d\theta/dP_{dark}$ through a reduction of $\tau_{qp}^*$ (or $Q$), as can be seen from Eq.~\ref{Eqn:response}. This degradation would lead to an increase in NEP. We therefore measured $\tau_{qp}^*$, $d\theta/dP_{dark}$, and the dark NEP before and after the proton beam irradiation. For an accurate determination of $\tau_{qp}^*$ and $d\theta/dP_{dark}$, measurements at various bath temperatures were required~\cite{janssen_equivalence_2014}.

Once $V$, $N_{qp}$, $\tau_{qp}^*$ and $\Delta$ are well determined, the dark NEP can be used as a proxy for a direct NEP measurement using a calibrated radiation source, typically referred to as the optical NEP \cite{janssen_equivalence_2014}.
In the experiment described in this work, we mainly use the dark NEP, as a measurement of the optical NEP requires a setup too elaborate to be installed in DRACuLA.

\subsection{Experimental setup and expected radiation dose}

\begin{figure*}
    \includegraphics[width = 0.9\textwidth]{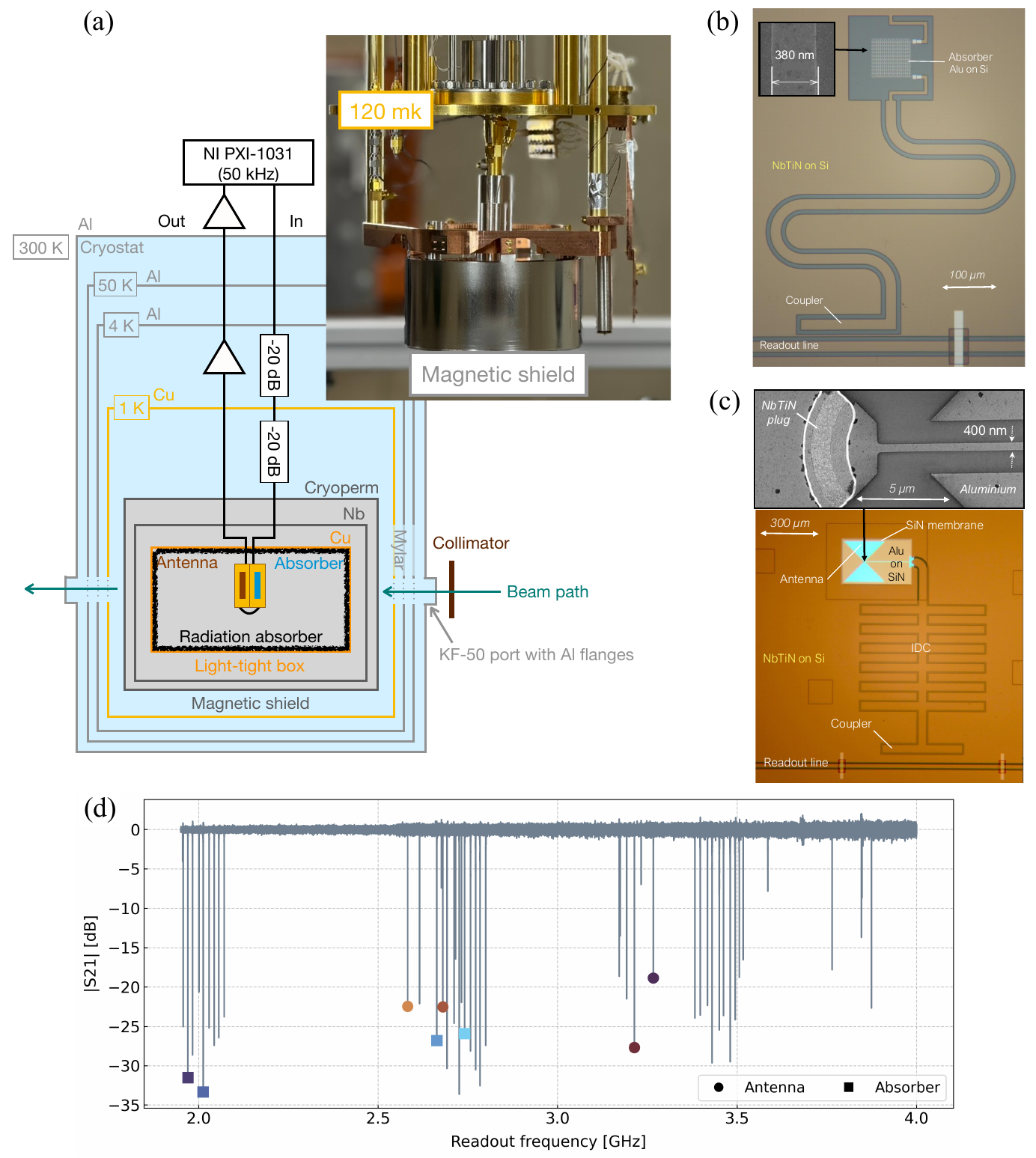}
    \caption{This figure presents the different elements of the experimental set-up: \textbf{(a)} block-diagram describing the experimental set-up and the electronics, with a picture of the 120 mK stage. \textbf{(b)} geometry of the absorber-coupled KIDs, figure adapted from Ref.~\citeonline{dabironezare2025}. \textbf{(c)} geometry of the antenna-coupled KIDs, figure adapted from Ref.~\citeonline{Baselmans_2022}. \textbf{(d)} full frequency-sweep scan of the 2-4 GHz bandwidth with indication of measured KIDs. The circle dots in brown shades are measured KIDs from the antenna array, and the square dots in blue shades are measured KIDs from the absorber array.}
    \label{fig:setup}
\end{figure*}

Figure~\ref{fig:setup} (a) shows a schematic drawing of the setup inside the cryostat. 
Two KID arrays were placed in copper (Cu) sample holders and mounted back-to-back. The sample holders were equipped with radiation absorbers, made from carbon loaded epoxy and silicon carbide (SiC) grains, and placed in a Cu light-tight box to prevent any stray light\cite{Baselmans_2012}. This assembly was placed inside a Niobium (Nb) and a Cryoperm magnetic shield, which provide magnetic shielding. This whole assembly was mounted on a cold finger connected to the mixing plate of the cooler, inside a 1K radiation shield. Coaxial cables were used to connect both KID arrays in series. The outbound line was connected with two low-noise amplifiers at the 4 K stage and outside of the the cryostat.

The KID arrays were aligned with the cryostat's outer window, made of aluminum, which was later aligned with the proton beam. Thin infrared filters made from 100~µm metallized mylar with Al coating were placed on the 50 K, 4 K and 1 K shields' windows to reduce the thermal load on the 120 mK plate while allowing passage for protons. A collimator was placed outside the cryostat, between the KF-50 port and the beamline, reducing the beam size to $20\times20$~mm. The proton beam passes through all these layers to reach the detector chips. 
From a GEANT4 simulation~\cite{Geant4}, we estimated that the proton energy was reduced by these layers from 184 MeV to approximately 150 MeV.

When protons pass through the detector, a part of energy is deposited and absorbed in the detector, which is called radiation dose. There are two types of radiation dose: Total Ionising Dose (TID) and Dose Displacement Damage (DDD) or Total Non-Ionising Dose. The final concentration of defects (or displacements of atoms) depends only on the DDD. 
The mean energy deposition by ionization caused by the proton beam can be calculated using the energy spectrum of Linear Energy Transfer (LET) \cite{LET_NIST}. We estimated the mean energy deposition of $\sim 11$~MeV/cm in aluminum (per proton) from the 150~MeV protons. Considering the thickness of the aluminum film of KID arrays, this corresponded to $\sim 20$~eV energy deposition in each film.
For displacement damage, the mean energy deposition can be calculated using the energy spectrum of Non Ionising Energy Loss (NIEL) \cite{NIEL}. For our experiment, it was calculated as $\sim 5.7$~keV/cm in aluminum (per proton), meaning $\sim 0.01$~eV energy deposition in the film from this physics process.
Based on these expected energy depositions, and the expected proton beam flux of $\sim 2 \times 10^6$ protons/cm$^2$/s from the accelerator, we estimated the required beam time of 12~hours to simulate a total dose equivalent to a 10 year mission at L2.

\section{Measurements}
The KID arrays were irradiated with a proton beam for about 12~hours, resulting in a beam fluence of $\sim 8.4 \times 10^{10}$~protons/cm$^{2}$ in total.
As a result, the TID applied to the aluminum film in this experiment was roughly $3.5 \times 10^{11}$~MeV/g, while the DDD was $1.8 \times 10^8$~MeV/g, or equivalently, the total radiation dose of 5.7~krad by adding the TID and the DDD~\footnote{Particles other than protons, such as gamma rays, electrons, and positrons from the decay of radio-activated isotopes within the setup, may also deposit energy in the detector. These particles can contribute to the TID, but not to the DDD. In the dose estimation, their contribution was neglected due to the difficulty of accurately modeling radio-activation effects. The reported total dose of 5.7~krad therefore represents a lower limit.}. Note that the TID and DDD to the Si substrate were almost the same as for the aluminum film due to their similar mass density, atomic number and atomic mass, while the total deposit energy on the Si substrate was much larger due to its thickness.

We measured $\tau_{qp}^*$, $d\theta/dP_{dark}$, and dark NEP before and after beam irradiation (refered as pre- and post-radiation hereafter).
All the measurements were carried out during a single cooldown in DRACuLA in dark conditions, and results presented here were carried out using our “single-tone” homodyne readout electronics~\cite{deVisserThesis2014, PDay_2003}.
Here, “single tone” means that only one excitation frequency can be transmitted throught the KID arrays and received at a time. This setup can acquire time ordered data (TOD) at a high sampling rate, here 50~kHz, which allows us to measure $\tau_{qp}^*$~\cite{Kaplan}.
However, since only one KID can be read out at a time, measuring $\tau_{qp}^*$ for all KIDs would have required a long measurement time that was not possible within the limited available time for the accelerator campaign.
We therefore selected eight representative KIDs, as indicated in Figure~\ref{fig:setup} (d), for full characterization.
The selection of the measured KIDs was driven by two factors. Firstly, some resonances had already been characterized in advance in a laboratory~\cite{Baselmans_2022}. Secondly, their frequencies were sufficiently spaced to avoid overlap between tones.
A set of pre-radiation measurements was performed to establish a consistent reference for comparison with post-radiation data in DRACuLA.
Resonance frequencies were identified through a frequency sweep, and the optimal power was determined by measuring the noise at different power levels.
With the optimal power, we measured both a frequency sweep and the noise at a various temperatures in a range of 120 - 280~mK, for an accurate determination of $\tau_{qp}^*$ and $d\theta/dP_{dark}$. 

During the pre-radiation measurements, we observed some transient events, here we call them “glitches", in the TOD. These events were due to the radiative environment in the experimental room at PARTREC. Gamma rays or particles (such as neutrons) from the radiative environment could reach and hit the KID arrays inside the cryostat, creating high-energy phonons propagating through the arrays and causing glitches.

For the measurements right after the irradiation, we observed an extremely high concentration of glitches in the TOD, making the baseline indistinguishable from the noise.
These glitches were mainly due to $\beta^{\pm}$ particles emitted from decays of $^{64}$Cu isotopes, which were created by the neutron capture process by Cu in the light-tight box during the irradiation.
The half-life time of $^{64}$Cu is 12.7~hours while the expected mean energy deposition of the emitted $\beta^{\pm}$ in the KIDs aluminum film is $20-30$~eV.
Therefore, the post-radiation dataset was acquired 24 hours after a total dose of 5.7~krad to reduce the number of glitches in the TOD.

\section{Results}

\begin{figure*}
    \centering
    \includegraphics[width=\textwidth]{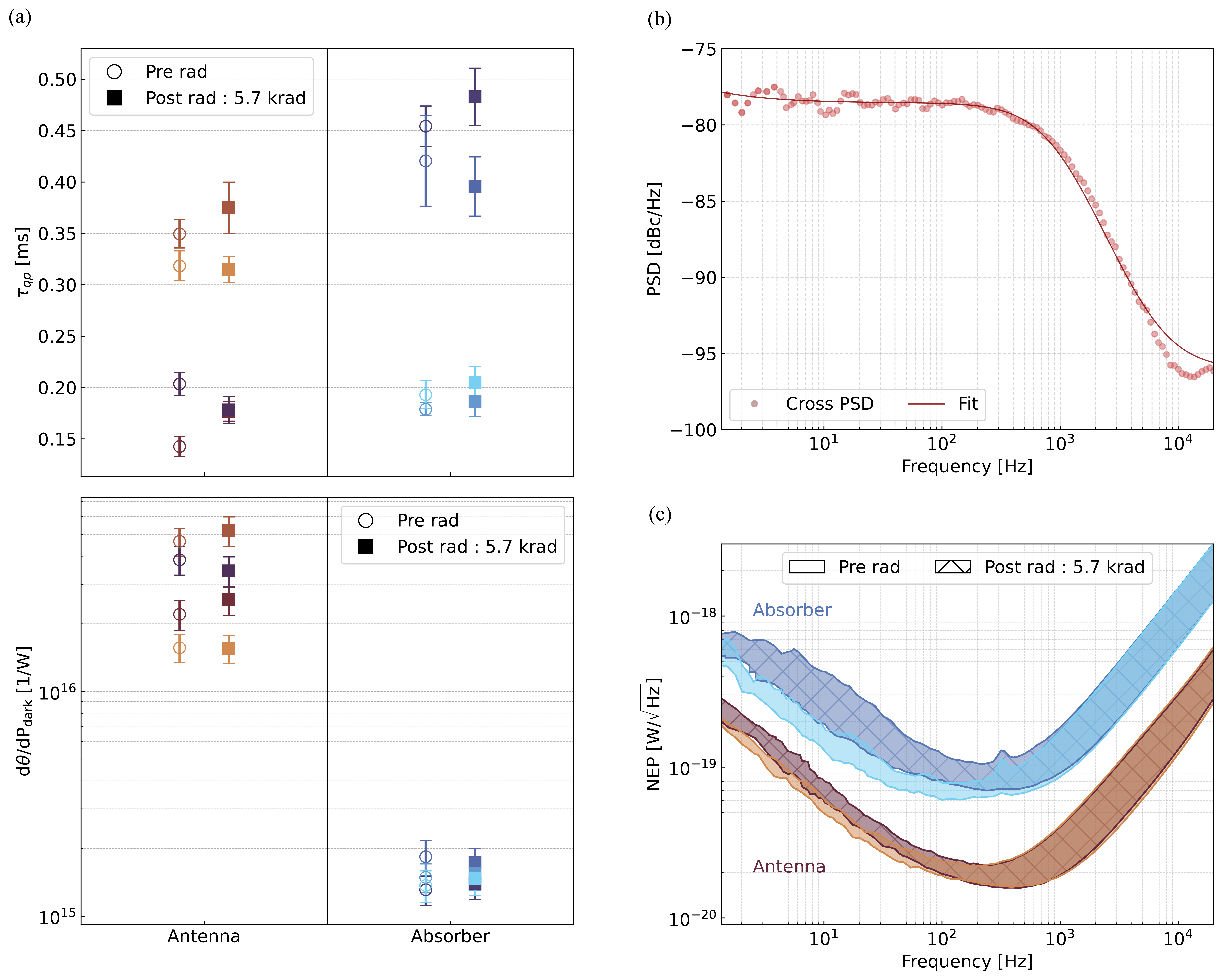}
    
\caption{This figure gather the results of the present work: The four antenna-coupled KIDs are represented in brown shades, while the four absorber-coupled KIDs are in blue shades. \textbf{(a)} Pre-(empty circles) and post-radiation (filled squares) results with uncertainties are compared for each KID, using consistent shade and color coding. \textit{Top:} quasi-particle lifetime ($\tau_{qp}^*$) values before and after 5.7~krad irradiation. Note that there are two groups of lifetime in both absorber and antenna arrays. They have different lifetime values by design as different geometries of absorber or antenna were incorporated into the irradiated arrays. \textit{Bottom:} Dark responsivity ($d\theta/dP_{dark}$) before and after 5.7~krad irradiation. \textbf{(b)} Phase noise PSD ($S_\theta$) and cross-PSD of one of measured KIDs at 200~mK, together with the fit of the roll-off (solid lines). \textbf{(c)} Comparison of the dark NEP for the absorber KIDs batch (blue) and the antenna KIDs batch (brown) before irradiation and after 5.7~krad irradiation (hatched areas).}
\label{fig:results}
\end{figure*}

Due to many glitches in the TOD in both pre- and post-radiation data, it was necessary to deglitch the TOD for proper estimation of noise PSD. To flag glitches in the data, the following procedure was applied: 
\begin{enumerate}
    \item After taking the rolling mean of the time stream data with a window size of 400~${\mu \text{s}}$, which corresponds to roughly $\tau _{qp}^*$, the standard deviation ($\sigma$) of the time stream data was calculated.
    \item Data points above $ 4 \cdot \sigma $ were removed, and the standard deviation of the resulting time stream data was calculated again.
    \item Using the newly calculated standard deviation with the same threshold as before, any remaining glitches were flagged again, and at the same time we also flagged with a $\sim$~2.2~ms window ($4-5 \cdot\tau _{qp}^*$) around each glitch to remove any remnants.
    \item These steps were repeated three times. As the number of deglitched points remained consistent between the second and third iterations, it was confirmed that three iterations were sufficient to flag all glitches. The noise PSD was then calculated on the timestream with the flagged data removed.
\end{enumerate}
Approximately 1$\%$ of the pre-radiation measurements were affected by glitches while about 15$\%$ of the data points were impacted in the post-radiation dataset.

The dark NEP was evaluated following Ref.~\citeonline{janssen_equivalence_2014} and Eq.~\ref{Eqn:NEP}. We measured both a frequency sweep and the noise over a range of bath temperatures, from which we evaluated $\tau_{qp}^*$ by following Eq.~2 in Ref.~\citeonline{de_Visser_2011}, and then calculated $d\theta/dN_{qp}$ by following Eq.~1 in Ref.~\citeonline{janssen_equivalence_2014}. 
We used pair-breaking efficiencies ($\eta_{pb}$) of 0.35 for the antenna coupled KIDs~\cite{Baselmans_2022} and 0.25 for the absorber coupled KIDs~\cite{Guruswamy_2014} for the calculation of $d\theta/dP_{dark}$ as shown in Eq.~\ref{Eqn:response}.
Then the dark NEP was calculated by following Eq.~\ref{Eqn:NEP}.

We evaluated $\tau_{qp}^*$ by fitting the roll-off of the cross-PSD, which was calculated from the cross-correlation of the phase and amplitude TOD of KIDs~\cite{deVisser2012APL}.
Since the generation and recombination noise is suppressed into the readout noise level by trapping at lower temperature than $\sim$200~mK~\cite{RooijPRL2025}, obtaining $\tau_{qp}^*$ in this way is difficult at lower temperatures.
To check $\tau_{qp}^*$ at the operation temperature of 120~mK, we fit the decay tails of glitches with an exponential function, and this method gave consistent results with $\tau_{qp}^*$ from the cross-PSD.
Figure~\ref{fig:results} (b) shows an example of the cross-PSD of one of KIDs at 200~mK with the fit of the roll-off.

The obtained $\tau_{qp}^*$ and $d\theta/dP_{dark}$ are presented in Figure~\ref{fig:results} (a) top and bottom, respectively, for absorber and antenna KIDs before irradiation (empty circles) and after the 5.7~krad dose (filled squares). For both $\tau_{qp}^*$ and $d\theta/dP_{dark}$, the post-radiation results align with the pre-radiation results within the measurement uncertainty.

Figure~\ref{fig:results} (c) presents the dark NEP for both type of KIDs, absorbers in blue and antenna in brown, for the pre-irradiation (solid-filled) and post irradiation (hatched). We observe an increase in dark NEP after irradiation which shows a clear frequency dependence and differs between absorber-coupled and antenna-coupled KIDs. At 10 Hz, the dark NEP increases by 53$\%$–207$\%$ for absorber-coupled detectors, while the increase for antenna-coupled detectors is significantly lower, ranging between 28$\%$ and 49$\%$.

\section{Discussion}
The observed rise in NEP after radiation is caused by an increase in noise, since both $\tau_{qp}^*$ and $d\theta/dP_{dark}$ do not change (see Eq.~\ref{Eqn:NEP}). This excess noise is possibly due to incomplete deglitching in the post-radiation data, which is especially hard for the absorber-coupled KIDs which are more sensitive due to being on the bare substrate~\cite{Karatsu_2019}. 

\begin{figure}
    \centering
    \includegraphics[width=0.9\linewidth]{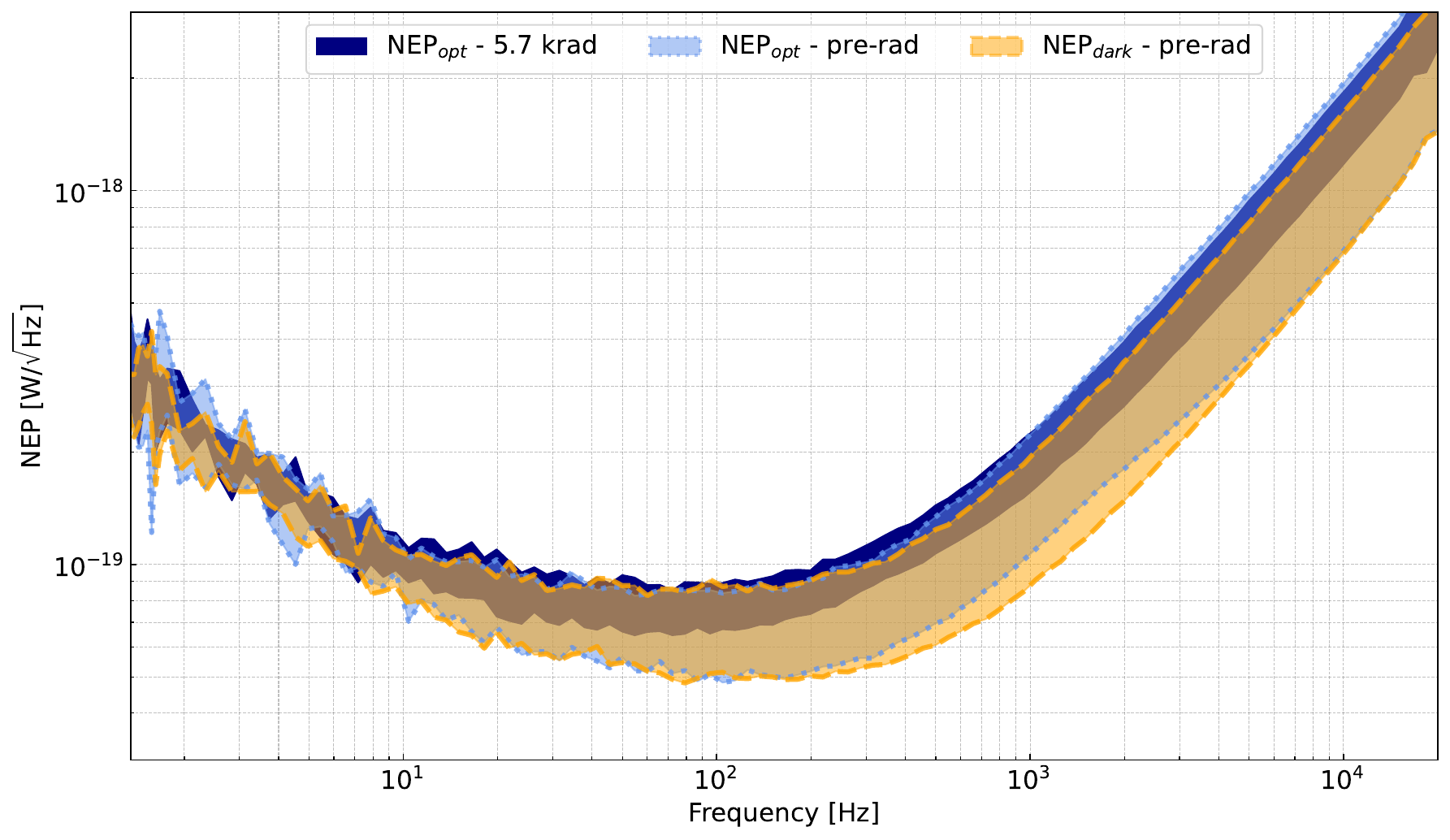}
    \caption{Comparison between the NEP measurements performed in a laboratory with ADR, before and after the total dose radiation, for the absorber KID array. The post-radiation optical NEP is presented in dark blue, the pre-radiation optical NEP in light blue (dotted-filled) and the pre-radiation dark NEP in orange (dashed-filled).}
    \label{fig:comparison_nep}
\end{figure}

To confirm the hypothesis of incomplete TOD deglitching, post-radiation measurements were performed again in a laboratory with the ADR setup for the absorber-coupled KID array. The results, all obtained using the same setup, are presented in Figure~\ref{fig:comparison_nep}, showing similar results among the pre-radiation optical NEP, pre-radiation dark NEP, and the post-radiation optical NEP. The good agreement between the optical and dark NEP supports our methodology to use the dark NEP as a proxi for the optical NEP.  Ultimately, no significant difference is observed in optical NEP between the pre- and post-radiation.
This result suggests that the degradation in NEP observed with DRACuLA at PARTREC was caused by the very high radiogenic environment and a partial incomplete deglitching.

These results confirmed that the proton beam irradiation did not cause any damage on the KID arrays. However, this raises the need to understand the underlying physics of the glitches in order to find an efficient way to deglitch data acquired at the L2 point.
More study of the subject is currently in progress and will lead to future publication.
Note that the glitch rate recorded for the post-radiation data in DRACuLA is 15--20~/s/cm$^2$, which is three to four times higher than the estimated glitch rate of $\sim 5$~/s/cm$^2$ while operating at L2 point~\cite{Planck_2014}. 

\section{Conclusion}
KIDs will be used for PRIMA, and it is one of the most promising detector technologies for future space missions observing in wavelengths from the millimeter to the far-infrared range.
We studied possible irradiation effects on KIDs in a cryogenic environment using a proton beam with an energy of 184~MeV at PARTREC.
Two types of KID array were cooled down to 120 mK: absorber-coupled and antenna-coupled KIDs. They were then irradiated with a total dose of 5.7~krad, equivalent to approximately 10 years of operation at the L2 Lagrange point. 
Comparing the pre- and post-radiation measurements revealed no significant change in the quasi-particle lifetime, $\tau_{qp}^*$, and the dark responsivity, $d\theta/dP_{dark}$.
The observed increase in the noise and NEP is attributed to incomplete deglitching due to a large number of glitches in the TOD. 
Subsequent measurements in a laboratory prove that the applied radiation dose does not cause any permanent damage to the irradiated KID arrays.

\subsection*{Disclosures}
The authors declare that there are no financial interests, commercial affiliations, or other potential conflicts of interest that could have influenced the objectivity of this research or the writing of this paper.

\subsection* {Code, Data, and Materials Availability} 
The processed data and code used for generating the figures in this paper are available on Zenodo: \url{https://doi.org/10.5281/zenodo.24496903}. The raw, unprocessed data are about 165 GB in size and are not publicly available due to their large size. The raw data are available upon request from the corresponding author.

\subsection* {Acknowledgments}
The authors acknowledge financial support from the Centre national d’études spatiales (CNES), France (ROR: https://ror.org/04h1h0y33). The DRACuLA facility has been co-funded by the Ile de France region through the DIM-ACAV program, the Paris-Saclay University, and the OSUPS (Observatoire des Sciences de l’Univers Paris-Saclay). The authors also thank the PARTREC staff from Gröningen University, Netherlands. Language models were used for language and grammar clean-up.


\bibliography{report}   
\bibliographystyle{spiejour}   


\vspace{2ex}\noindent \textbf{Anaïs Besnard} is a Ph.D. student in Instrumental Cosmology at the Institut d'Astrophysique Spatiale, Orsay, France. She is a member of the LiteBIRD, BISOU, and FOSSIL collaborations. Her research aims to improve the instrumental sensitivity of future space missions. She studies the impact of cosmic rays on prototypes of high-sensitivity detectors, TES and KID technologies, through irradiation campaigns carried out for LiteBIRD and PRIMA respectively. Her work seeks to understand and mitigate these effects in order to ensure unbiased measurements in a space environment. She is also involved in the development of a Closed-Cycle Dilution Refrigerator (CCDR) and its subsystems.

\vspace{1ex}
\noindent \textbf{Giulia Conenna} has been a postdoctoral researcher at SRON (Space Research Organization of the Netherlands) since November 2024, after receiving her PhD from the University of Milano-Bicocca with a thesis in experimental cosmology. At SRON, she is part of the PRIMA and POEMM collaborations, actively contributing to both experimental work and data analysis, including improvements in the sensitivity of Kinetic Inductance Detectors (KIDs). With her background in cosmology, she is also involved in the FOSSIL experiment, and at SRON she contributes to the development of low-Tc KIDs.

\vspace{1ex}
\noindent \textbf{Valentin Sauvage} is a postdoctoral researcher in Instrumental Cosmology. He works on various research topics such as the development of the space Closed-Cycle Dilution Refrigerator, the cryogenic development of the balloon-borne mission BISOU, and the study of the impact of cosmic rays on future detectors for the LiteBIRD mission (JAXA). He is a main contributor of the FOSSIL space mission (ESA M8 proposal) on the thermal architecture of the instrument.

\vspace{1ex}
\noindent \textbf{Kenichi Karatsu} is an Instrument Scientist with SRON. He received his Ph.D. degree in science in 2011 from Kyoto University with study of proton spin structure at the RHIC-PHENIX experiment. In 2015, He joined the DESHIMA project at TU Delft/SRON. His main role is to lead laboratory evaluation and telescope campaign of the DESHIMA instrument to the ASTE telescope. His research interest is to develop an experimental instrument for revealing mysteries of the Universe.

\vspace{1ex}
\noindent \textbf{Stephen Yates} is an instrument scientist at SRON, the Space Research Organisation of the Netherlands (since 2006). He received a Ph.D. from the University of Bristol (2003), followed by work at the CNRS-CRTBT (now Institut Neél) Grenoble on experimental low temperature physics (2004) and astronomical instrumentation (2004-2006). His current interests concentrate on Kinetic Inductance Detector development for sub-mm to far infra-red astronomical applications but also include a wider interest in device physics and superconductivity, optics, and full end to end instrument characterization and performance. Dr. Yates has published over 90 papers.

\vspace{1ex}
\noindent \textbf{Jochem Baselmans} is professor in experimental astronomy at the Delft University of Technology, senior instrument scientist at the SRON Space Research Organisation Netherlands and member of the global faculty of the University of Cologne. He graduated in 1998 and received his Ph.D. in 2002 both at the University of Groningen, studying superconductivity in normal metal Josephson junctions. He is an expert in Kinetic Inductance Detectors (KIDs) and systems. He has pushed the sensitivity and TRL of KID systems to a level suitable for operation in cryogenically cooled space-based observatories. He is CO-I of PRIMA, as the lead for the detector development for its European instrument PRIMAger, detector lead for the POEMM balloon project and the 18k pixel AMKID camera on the APEX telescope. Furthermore, he pioneered the concept of the on-chip spectrometer and is the lead system engineer for the first on-chip spectrometer instruments, DESHIMA (2017) and DESHIMA 2.0 (2023-2024).

\vspace{1ex}
\noindent \textbf{Bruno Maffei} has worked in the field of Radio to Sub-millimetre wave astronomical instrumentation since 1990, when I started my PhD at the IAS-Orsay, France. After a short stay at CalTech and IPAC in Pasadena, he has then worked for 21 years in the UK (QMUL, Cardiff University and The University of Manchester / Jodrell Bank Centre for Astrophysics), contributing to the development of several astronomical instruments, such as the Planck-HFI and Herschel-SPIRE ESA missions. Since 2016, he is working on several R\&D programmes on cryogenics, optics and detector testing, leading an instrumentation group within the Cosmology team at the IAS to support developments for CMB and far-infrared Astronomy. He his now the P.I. of the BISOU balloon-borne project as well as one of the main contributors of the FOSSIL mission proposal to ESA.

\vspace{1ex}
\noindent \textbf{Lorenza Ferrari} has 15+ years' experience in infrared instrumentation for ground- and space-based missions, with expertise in cryogenic detector development, assembly, integration, and verification. She has managed complex, multi-disciplinary projects, including acceptance and testing of PLATO flight cameras, development of KID imaging instruments deployed at APEX, and current management of SRON deliverables for NASA’s PRIMA and POEMM mission in Phase A.

\listoffigures
\listoftables

\end{spacing}
\end{document}